# Real-Time Machine Learning Based Fiber-Induced Nonlinearity Compensation in Energy-Efficient Coherent Optical Networks

**Elias Giacoumidis [1,*,†], Yi Lin [1,†], Michaela Blott[2], and Liam P. Barry[1]**

[1] Dublin City University, School of Electronic Engineering, Radio and Optical Communications Lab, Glasnevin 9, Dublin, Ireland.

[2] Xilinx Inc, 2020 Bianconi Ave, Citywest Business Campus, Dublin, D24 T683, Ireland.

* Correspondence: elias.giacoumidis@dcu.ie

† Equally contributed to this work.

**Abstract:** We experimentally demonstrate the worlds' first field-programmable gate-array (FPGA)-based real-time fiber nonlinearity compensator (NLC) using sparse K-means++ machine learning clustering in an energy-efficient 40-Gb/s 16-quadrature amplitude modulated (QAM) self-coherent optical system. Our real-time NLC shows up to 3 dB improvement in Q-factor compared to linear equalization at 50 km of transmission.

## 1. Introduction

Surging data traffic demands caused by bandwidth-hungry Internet services such as video streaming or cloud computing pose a significant challenge to the underlying fiber-optic communication systems. The only viable solution for increasing the data rates is the employment of advanced modulation formats (e.g. 16-QAM). The core difficulty on the transmission of such signals is the Kerr-induced fiber nonlinearity [1] which is responsible for nonlinear optical effects such as FWM [2]. The optical Kerr effect is attributed to the so-called nonlinear Shannon capacity limit [2] which sets an upper bound on the achievable data rate in optical fiber communications when using traditional linear transmission techniques. There have been extensive efforts in attempting to surpass the nonlinear Shannon limit through several fiber nonlinearity compensators (NLCs) [1-5], with the most well-known being the optical phase conjugation [1], digital back-propagation [2, 3], phase-conjugated twin-waves [4], and the nonlinear Fourier transform (NFT) [5]. However, besides these solutions being either complex [1-3, 5] or spectrally inefficient [4], more importantly, they traditionally compensate deterministic nonlinearities thus ignoring critical stochastic nonlinear effects such as the interplay between amplified spontaneous emission from optical amplification with fiber nonlinearity. Recently, digital machine learning has been under the spotlight for compensation of nonlinear distortions, harnessing various algorithms that perform classification such as supervised artificial neural networks [6] and support vector machines [7] or unsupervised clustering such as K-means, fuzzy-logic C-means, hierarchical [8, 9] and affinity propagation [10]. Digital-based machine learning digital blocks are independent from mathematically tractable models and can be optimized for a specific hardware configuration and channel [11]. Hitherto, reported machine learning based NLCs have been implemented offline using conventional software platforms (e.g. Matlab®). However, there is a tremendous need for a practical algorithm for application in real-time communication links that can maximize transmission performance without sacrificing network latency and energy efficiency. Recent advances in the field of field-programmable gate-arrays (FPGAs) have made it possible to process high bandwidth digital signals in next-generation telecommunication systems. On the other hand, unsupervised machine learning clustering is more attractive than supervised algorithms because training signals that limit signal capacity and increase computational time are absent.

In this work, we experimentally demonstrate the first FPGA-based NLC using sparse K-means++ clustering for 40-Gb/s 16-QAM energy-efficient self-coherent systems [12] (see Fig. 1). We show that our proposed NLC can offer up to 3 dB Q-factor enhancement for transmission at 50 km of SSMF.

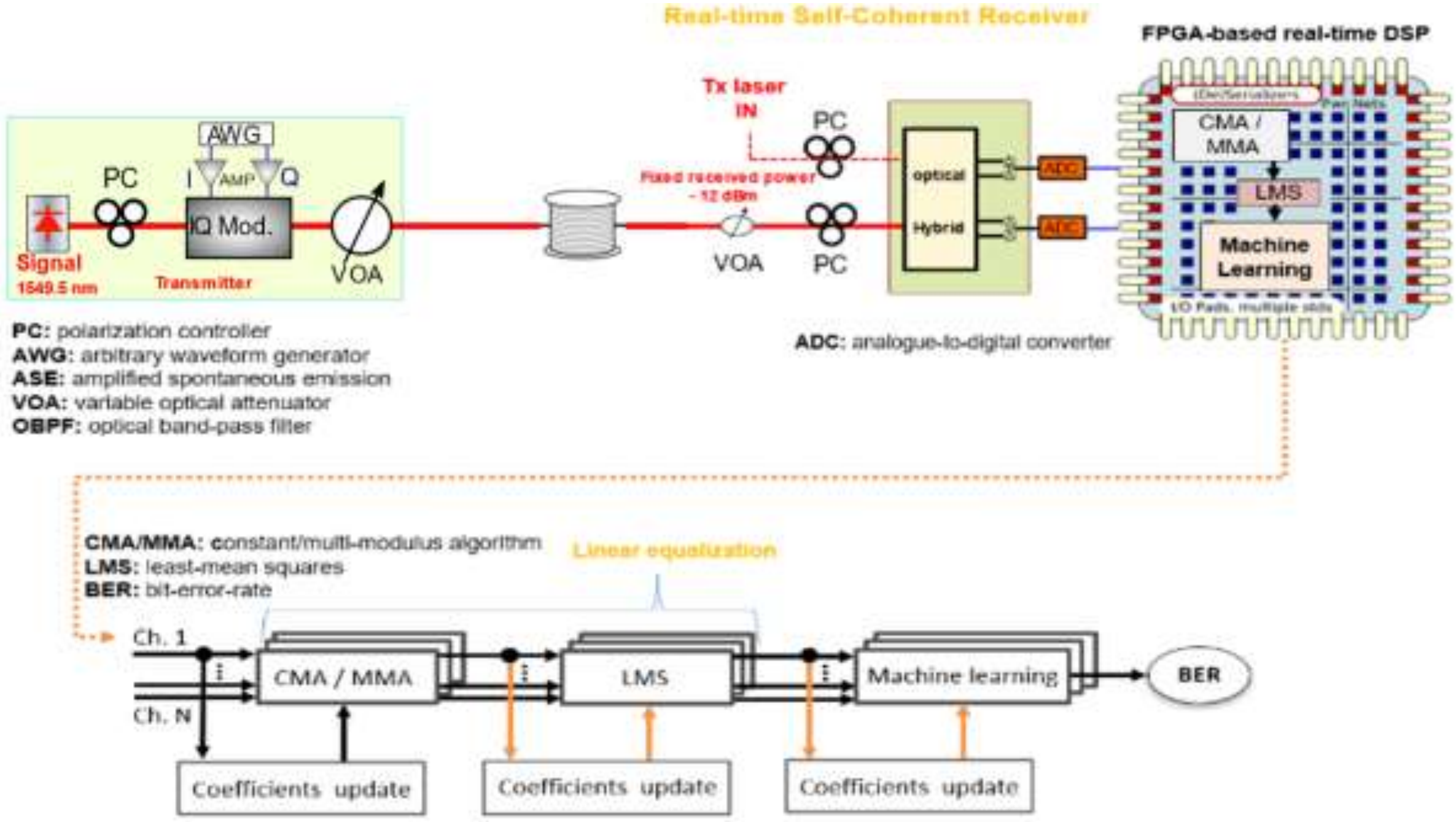

**Figure 1.** Self-coherent system for 50 km transmission incorporating the machine learning based FPGA receiver.

## 2. FPGA design and experimental setup

K-means clustering is a method of vector quantization that has been popular for cluster analysis in data mining. It aims to partition $n$ observations into $K$ clusters in which each observation belongs to the cluster with the nearest mean, serving as a prototype of the cluster. However, K-means is computationally inefficient especially for real-time signal processing where many calculations are required for thousands or millions of data points in a very limited timeframe (msec). Here, we propose a sparse K-means++ based NLC for real-time self-coherent optical signals. In this algorithm, we take regular time intervals (sparsity) from the real-time incoming signal to update the centers of the clusters thus eliminating time delays and latency. The clustering algorithm is based on the Lloyd's approach [8, 9] and its objective is to minimize the total intra-cluster variance, or, the squared error function: $J = \sum_{j=1}^{K}\sum_{i=1}^{n}\left\|x_i^{(j)} - c_j\right\|^2$. Where $J$, $K$, $n$, are the objective function, number of clusters, and number of symbols, respectively. $x_i$ is the $i^{th}$ symbol and $c_j$ is the centroid for a cluster $j$. The inside term in the norm $\|.\|$ defines the distance function between a symbol and a centroid. In the algorithm, we have knowledge about the centers (i.e. the '++' term) of the adopted modulation format level to enable faster convergence and adaptivity.

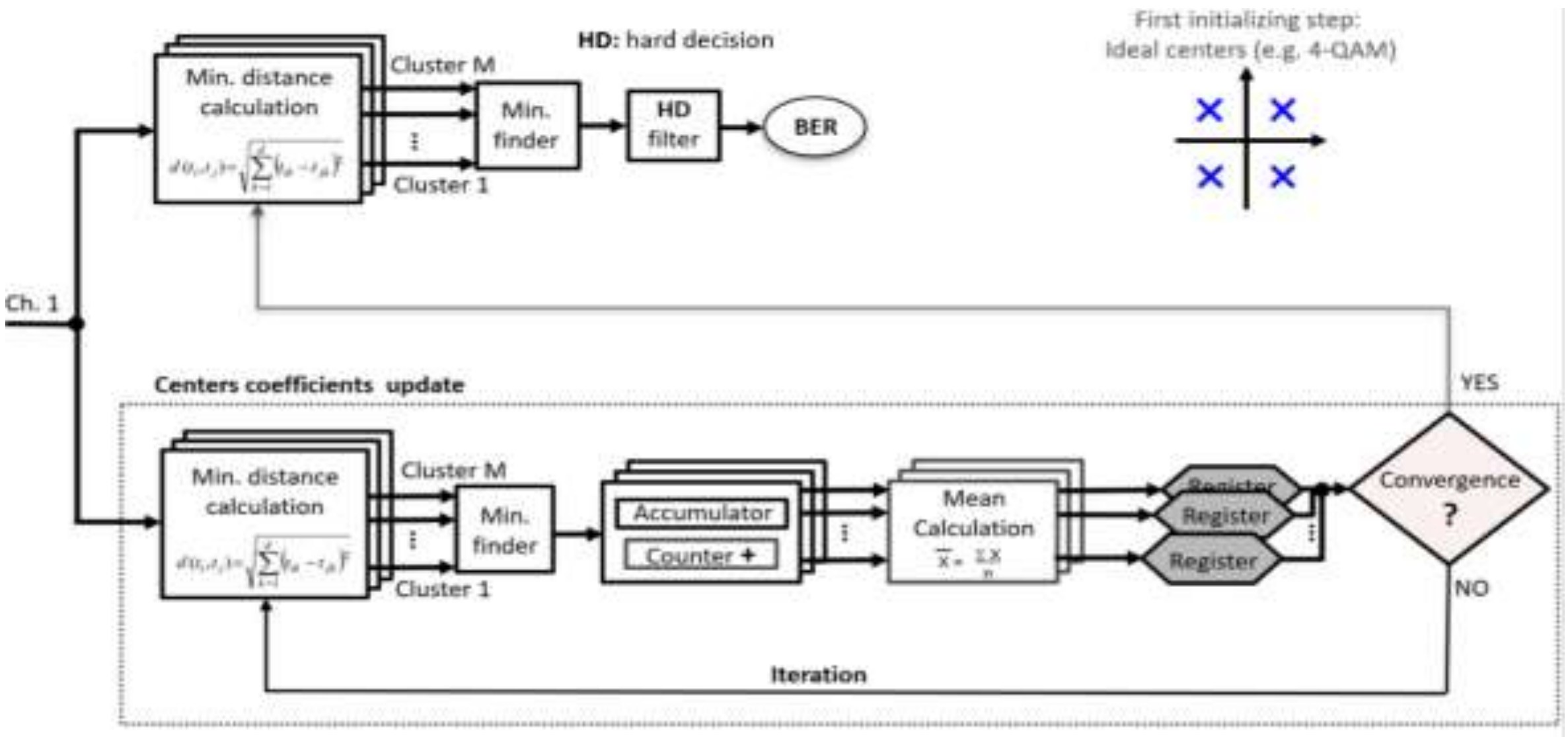

(a)

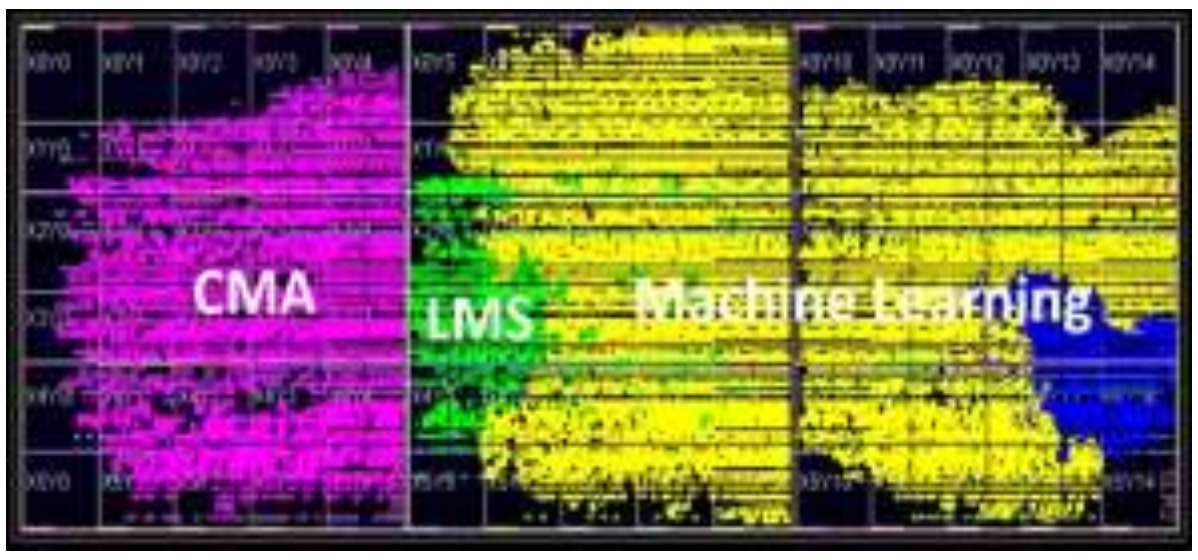

(b)

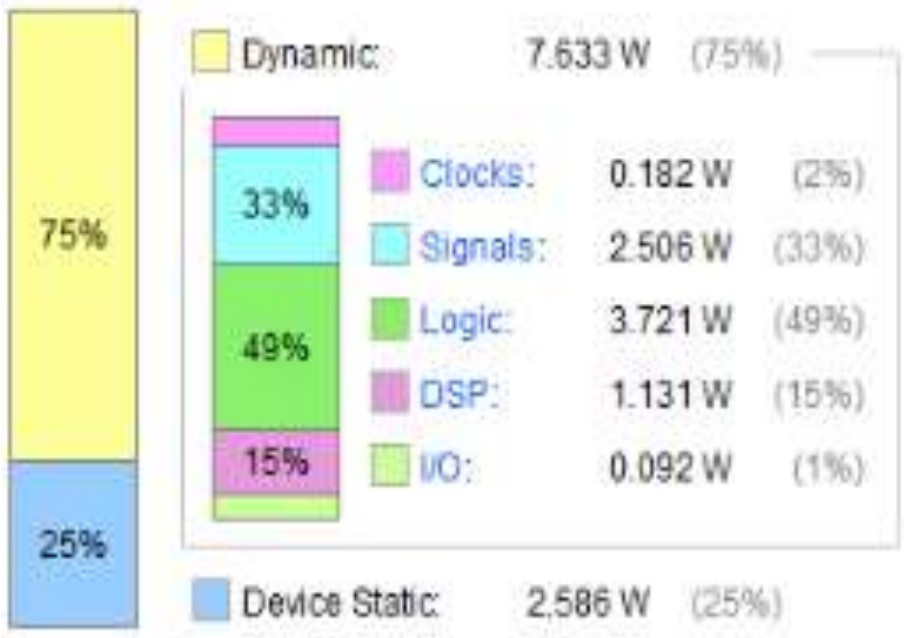

(c)

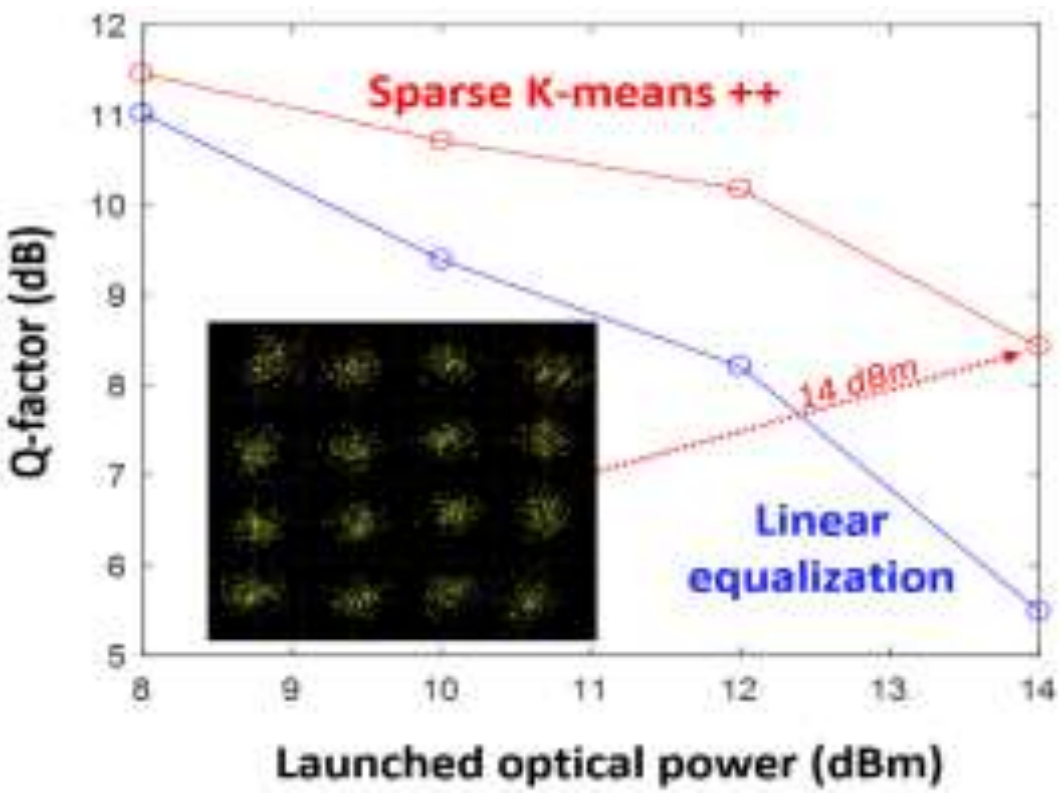

(d)

**Figure 2.** (a) Design of sparse K-means++ on FPGA. Inset: First initialization step with ideal centres in a 4-QAM signal. N: number of parallel FPGA channels; M: number of clusters in the constellation diagram. (b) Floorplan of the DSPs on the FPGA. CMA (purple); K-means hard-decision (yellow) and coefficients update (blue); LMS (green). (c) On-chip power report (power estimation from Xilinx Synthesized netlist). (d) Performance of sparse K-means ++ for 16-QAM at 50 km.

The steps of our algorithm are given below, while the conceptual diagram in Fig. 2(a) shows an example of the four centers' updates of a 4-QAM signal:

1. Initiate $K$ cluster centers (centroids) by averaging the received data per cluster after linear equalization;
2. Compute point-to-cluster-centroid distances of all observations to each centroid;

3. Assign each observation to the cluster with the closest centroid (Batch update);
4. Compute the average of the observations in each cluster to obtain $K$ new centroid locations;
5. Repeat steps 2 through 4 until cluster assignments do not change, or the maximum number of iterations is reached.

In our NLC the number of iterations was fixed, as further iterations didn't improve the Q-factor. In Fig. 2(a), we show the design of the NLC. As shown, in order to update the centers in K-means, only information from one of the N-parallel FPGA-channels (32 in total) is required for the I and Q components, thus significantly reducing the processing time. The system is split into two parallel processors, where in the upper diagram the minimum Euclidean distance among sparse received data per constellation cluster M and the ideal centres is first calculated (i.e. the '++' term). The inset of Fig. 2(a) shows an example of the first ideal centres initialization of a 4-QAM signal. Afterwards, hard decision is executed and the BER/Q-factor is calculated. After this initialization, the lower arm in the design is activated where we calculate the coefficients of the K-means following a similar procedure as before (i.e. via Euclidean minimum distance calculation per cluster) and estimate the mean for each center, and then update them in each iteration depending on renewed cluster mean. In the last iteration of the process, we re-assign symbols until all of them are properly assigned, considering the minimum BER value has been reached (convergence). The FPGA used in our experiment was a Xilinx Virtex Ultrascale+VCU118 Evaluation FPGA Platform. The self-coherent system (Fig. 1) employs a 10 GBaud 16-QAM signal at 40-Gb/s. The transmitter-DSP was processed offline in Matlab® using a look-up-table (LUT) in which pre-distortion was used to mitigate the opto-electronic components impairments similarly to [13]. A narrow linewidth (<100 kHz) ECL was tuned to 1549.5 nm and using an arbitrary waveform generator (AWG) operating at 20 GS/s, two uncorrelated pseudo-random level signals ($2^{15}$-1) were applied to the IQ modulator to generate the 16-QAM signal. After IQ modulation the optical signal was transmitted over 50 km of SSMF. At the receiver, the optical signal was converted to an electrical one using a self-homodyne coherent receiver. Afterwards, the signal was captured by a real-time oscilloscope sampled at 50 GS/s and directly after the ADCs the resultant electrical signal entered our FPGA board. First, the real-time signal passed through the constant modulus algorithm (CMA) combined with multi-modulus algorithm (MMA) for signal equalization. Afterwards, the signal went through a least-mean squares (LMS) digital filter for carrier phase recovery, and finally our proposed machine learning was processed in which hard and soft decision were included before BER/Q-factor (=$20\log_{10}[\sqrt{2}\,erfc^{-1}(2BER)]$) calculation. The design structure of the CMMA/MMA and LMS for real-time application is identical to that reported in [14]. For our FPGA design, 32 parallel channels (Ch. N) were attached to meet the required 20 GHz bandwidth considering the available clock rate per channel is 312.5 MHz. From Ch. 1 to Ch. N, we only selected one channel and used it to update the information of the centers in K-means. However, for each channel we processed separately the I and Q components.

**3. Results**

Fig. 2(b) and Table 1 present the resources utilization of the receiver DSP after FPGA synthesis and implementation, and Fig. 2(c) shows the on-chip power which is in total 10.219 W. It is evident that sparse K-means++, in which the hard-decision (yellow) and the coefficients update (blue) are included, covers > 60% of the total complexity of the DSP receiver (e.g. LUTs, configurable logic block (CLB), 8-bit carry chain per CLB (CARRY8)). The rest of the complexity is allocated to the CMA (purple) and the LMS (green). While this is contributed to a self-coherent system, it is worth noting that in a conventional coherent system, additional linear equalization is required for the frequency offset compensation between the transmitter and receiver laser. Finally, since our system was tested for short-each transmission at 50 km, chromatic dispersion compensation was ignored in the linear equalization process. We thus envisage that the total complexity of the linear equalizer in traditional coherent homodyne systems is much larger than the proposed machine learning algorithm.

**Table 1.** Key FPGA resources of developed DSP.

| DSP part | LUTs | Registers | CARRY8 | CLB |
|---|---|---|---|---|
| LMS | 26630 | 5940 | 3404 | 5655 |
| Sp. K-means++ | 431149 | 60316 | 32444 | 67973 |
| CMA | 164538 | 28336 | 11264 | 25970 |

Finally, in Fig. 3(d), we show the transmission performance of the 16-QAM 10 Gbaud signal at 50 km when using linear equalization and sparse K-means++ for launched optical powers (LOPs) of up to 14 dBm. We show that our NLC enhances the Q-factor by 3 dB at 14 dBm of LOP (a constellation diagram is also included). This is attributed to the compensation of self-phase modulation as a single-channel/carrier is transmitted.

## 4. Conclusion

We experimentally demonstrated a novel, practical fiber-induced nonlinearity compensator, the sparse K-means++, for 40-Gb/s real-time 16-QAM energy-efficient self-coherent systems. At 50 km of commercial fiber transmission, our approach improved the Q-factor by up to 3 dB over linear equalization. This is the worlds' first implementation of machine learning in an FPGA board for compensation of nonlinearities in optical networks. We believe our technique can be also valuable for other modulation techniques such as fast optical orthogonal frequency division multiplexing [15-17].

## 5. Acknowledgments

This work was supported by SFI through 13/RC/2077, 12/RC/2276, 15/US-C2C/I3132, the HEA INSPIRE, and the EU/EDGE Marie-Curie COFUND programme (713567).